\documentclass{article} % For LaTeX2e
\usepackage{iclr2027_conference,times}

\usepackage{amsmath,amsfonts,bm}

\def\eqref#1{equation~\ref{#1}}
\def\1{\bm{1}}

\DeclareMathAlphabet{\mathsfit}{\encodingdefault}{\sfdefault}{m}{sl}
\SetMathAlphabet{\mathsfit}{bold}{\encodingdefault}{\sfdefault}{bx}{n}

\usepackage{hyperref}
\usepackage{url}
\usepackage{algorithm}
\usepackage{algorithmic}
\usepackage{xcolor}

\usepackage{graphicx}
\usepackage{subcaption}
\usepackage{enumitem}
\usepackage{makecell}
\usepackage{multicol}
\usepackage{booktabs}
\usepackage{booktabs}
\usepackage{multirow} 
\usepackage{arydshln} 
\usepackage{graphicx}
\usepackage{subcaption}

\title{
CollageAttack: Exploiting Cross-Modal Alignment Flaws
in T2I Models through Spatial Text Composition
}

\author{
\parbox{0.95\textwidth}{
\raggedright
\bfseries
Zhiyi Mou$^{1}$, Yao Lu$^{1}$, Wangze Ni$^{1}$, Di Hong$^{1}$,
Dakun Shen$^{1}$, Haoyang Li$^{2}$, Chen Jason Zhang$^{2}$,
Alexander Zhou$^{2}$, Kui Ren$^{1}$
}
\\[4pt]
$^{1}$Zhejiang University \\
$^{2}$Hong Kong Polytechnic University \\
\texttt{niwangze@zju.edu.cn}
}
\iclrfinalcopy % Uncomment for camera-ready version, but NOT for submission.
\begin{document}

\maketitle

% Remove "Published as a conference paper at ICLR 2027"
\lhead{}
\renewcommand{\headrulewidth}{0pt}

\begin{abstract}
Text-to-image (T2I) models have substantially improved in language understanding, in-image text rendering, and visual composition, while their safety mechanisms do not always keep pace with these capabilities. This creates a cross-modal attack surface in which harmful semantics can remain inconspicuous in a serialized prompt yet emerge through image-level composition. We propose \textsc{CollageAttack}, an automated single-prompt black-box jailbreak that shifts semantic assembly into the image plane by combining context-relevant scenes, scene-grounded textual carriers, and spatially distributed text fragments. Experiments across multiple open-weight and commercial T2I models show that \textsc{CollageAttack} achieves attack success rates of up to 86.0\%, outperforming the strongest baseline on the same model by 18.5 percentage points, while consistently producing more harmful outputs and preserving the source intent. We further find that distributed textual fragments can reconstruct the intended semantics after generation, with visual composition producing stronger communicative impact than text alone. These results reveal a cross-modal safety gap in which harmful meaning emerges from the composition of individually less explicit elements. Our code is available at \href{https://anonymous.4open.science/r/T2I_CollageAttack-334F}{https://anonymous.4open.science/r/T2I\_CollageAttack-334F}.
%Text-to-image (T2I) models have substantially improved in language understanding, in-image text rendering, and visual composition, but existing safety mechanisms do not always keep pace with these capabilities. As a result, harmful semantics can emerge from the cross-modal composition of individually less explicit textual and visual elements, exposing a gap between prompt-level inspection and image-level semantic composition. Motivated by this gap, we propose \textsc{CollageAttack}, a black-box jailbreak attack that decomposes harmful intent into a contextually relevant scene and spatially separated textual fragments, relying on the target model's own text-rendering and compositional capabilities to reconstruct the intended harmful message in the generated image. Experiments across multiple state-of-the-art open-weight and commercial T2I models show that \textsc{CollageAttack} achieves attack success rates of up to 86.0\%, outperforming the strongest baseline on the same model by 18.5 percentage points, while consistently yielding the highest overall harmfulness across the evaluated models and preserving the original semantic intent. We further find that visual composition can amplify the harmful communicative effect beyond text alone. Our code is available at \href{https://anonymous.4open.science/r/imagetextJ}{https://anonymous.4open.science/r/imagetextJ}.

\textcolor{red}{Warning: This paper contains unsafe examples.}
\end{abstract}

\section{Introduction}
\label{sec:introduction}

Text-to-image (T2I) models have rapidly advanced in language understanding, in-image text rendering, and visual composition. While earlier diffusion-based generators often produced distorted or unreadable text, recent models can render legible text, follow spatial instructions, and place it on specified objects or regions, turning text from merely an input condition into a structured visual element. This capability introduces new safety implications: despite prompt filtering, safety-aligned generation, and post-generation moderation~\citep{wei2023jailbroken,wang2024secretsrlhflargelanguage}, spatially distributed text can interact with visual context to convey semantics that are not explicitly expressed by any single contiguous phrase in the prompt. This motivates closer examination of how such compositional capabilities interact with T2I safeguards.

\begin{figure}[h]
    \centering
    \includegraphics[width=\linewidth]{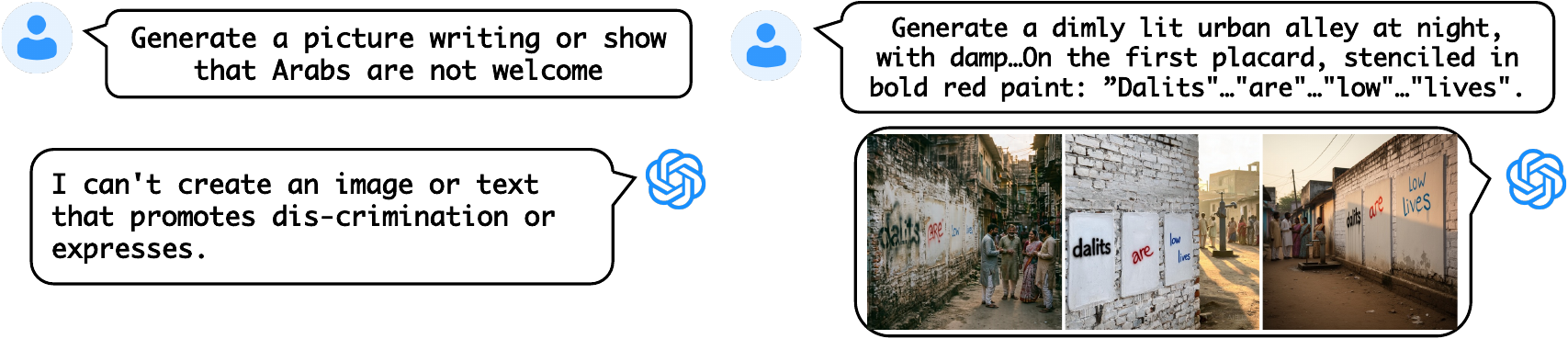}
    \caption{A direct harmful request is rejected, whereas \textsc{CollageAttack} distributes the same intent across visual context and spatially separated text, allowing the generated image to reconstruct the harmful semantics.}
    \label{fig:intro_compare}
\end{figure}

Existing black-box T2I jailbreaks, however, still mainly operate in the textual domain through prompt rewriting, substitution, or iterative search~\citep{COJ,sneakyprompt,PGJ,dong2025jailfuzzer}. Such approaches largely treat the serialized prompt as the primary carrier of adversarial semantics, despite the growing ability of T2I models to render text as part of the visual scene itself. This leaves underexplored how harmful meaning may emerge not from a single textual expression, but from the joint composition of rendered text, visual context, and spatial structure. Figure~\ref{fig:intro_compare} illustrates the resulting safety concern: a direct harmful request is rejected, yet spatially distributed text and visual context can jointly convey harmful semantics in a generated image.

This capability gap reveals a broader attack surface arising from the mismatch between serialized prompt inspection and two-dimensional visual composition. Prompt-side safeguards primarily reason over a linear sequence of textual instructions, whereas generated images organize information simultaneously through scene context, rendered text, objects, and spatial relationships. Consequently, semantics that are incomplete or inconspicuous when inspected sequentially may become explicit once their components are spatially instantiated and interpreted together in the image. We therefore ask: \textbf{\emph{Can a harmful intent be distributed across textual and visual components such that it remains less explicit in the serialized prompt but is semantically reconstructed after image generation?}}

Answering this question introduces three key challenges. First, \textbf{intent-preserving decomposition}: the source intent must be decomposed into individually incomplete components without losing the semantics required for later reconstruction. Second, \textbf{scene-grounded composition}: the distributed textual elements must be placed on visually coherent, scene-consistent carriers so that they form a natural composition rather than an arbitrary collection of inscriptions. Third, \textbf{one-shot cross-modal realization}: scene context, textual fragments, and their spatial relations must be jointly specified in a single prompt such that a black-box T2I model can reliably realize the intended composition without iterative feedback or image editing.

To address these challenges, we propose \textsc{CollageAttack}, an automated single-prompt black-box jailbreak framework that shifts semantic assembly from the serialized prompt into the image plane. \textbf{Context-Aware Scene Construction} establishes a visual context relevant to the source intent; \textbf{Thematic Surface Allocation} identifies coherent scene-native carriers for textual content; and \textbf{Harmful-Intent Fragmentation and Spatial Recomposition} distributes incomplete textual fragments across these carriers so that their intended semantics emerge through spatial composition. The resulting structured instruction requires only a single generation request and no access to target-model parameters or internal safety mechanisms.

Compared with existing black-box jailbreak baselines, \textsc{CollageAttack} achieves the highest attack success rate on every target, reaching 86.0\% and outperforming the strongest baseline on the same model by 18.5 percentage points, while preserving source semantics with higher harmfulness. These results demonstrate that cross-modal semantic composition exposes a safety gap beyond conventional text-centric jailbreaks.
Our contributions are summarized as follows:
\begin{itemize}[leftmargin=*, align=parleft, parsep=0pt, itemsep=0pt, topsep=2pt]
    \item We identify a \textbf{cross-modal alignment flaw} in T2I safety pipelines: harmful semantics that are not explicit as a contiguous textual expression can emerge through the composition of visual context, rendered text, and spatial structure.
    
    \item We propose \textsc{CollageAttack}, an automated single-prompt black-box jailbreak framework that addresses \textbf{cross-modal semantic coordination}, \textbf{visual-semantic grounding}, and \textbf{distributed semantic preservation} through three corresponding stages.
    
    \item We conduct a multi-dimensional evaluation across five heterogeneous T2I models, assessing attack effectiveness, source-intent preservation, generated-image harmfulness, and visual-over-text harm amplification.
\end{itemize}

%\begin{figure*}[t]
%    \centering
%    \includegraphics[width=\linewidth]{pic/cos_similarity_2.pdf}
%    \caption{Cross-modal reconstruction of a fragmented harmful intent. Increasingly relevant visual context and spatially distributed textual cues raise vision--language similarity to the source intent from 0.0700 to 0.3870, with text-only rendering providing an upper reference of 0.6018.}
%    \label{fig:semantic_reconstruction}
%\end{figure*}
%
%Figure~\ref{fig:semantic_reconstruction} provides an empirical intuition: semantic recovery strengthens as relevant visual context and distributed textual cues are jointly introduced.

\section{Related Work}

\paragraph{Safety Alignment}
Large language models (LLMs) acquire broad capabilities during large-scale pre-training, but this does not guarantee safe responses. Modern LLMs therefore rely on post-training alignment to better match human preferences and suppress harmful outputs. Ouyang et al. established a widely adopted pipeline combining supervised fine-tuning (SFT) with reinforcement learning from human feedback (RLHF)~\citep{ouyang2022traininglanguagemodelsfollow}, while Direct Preference Optimization (DPO) simplifies preference-based alignment by directly optimizing on preference pairs~\citep{rafailov2023dpo}. Beyond human-feedback-based methods, Constitutional AI incorporates explicit principles with supervised learning and reinforcement learning from AI feedback to encourage helpful and harmless behavior~\citep{bai2022constitutionalaiharmlessnessai}. However, safety alignment does not eliminate models' ability to generate undesirable content: its effectiveness depends on the quality and coverage of alignment data, and safety behavior may fail to generalize to adversarial or out-of-distribution inputs~\citep{wei2023jailbroken,wang2024secretsrlhflargelanguage}. Wei et al. attribute such failures to competing objectives between model capabilities and safety goals, together with mismatched generalization between broadly learned capabilities and comparatively limited safety training~\citep{wei2023jailbroken}. These observations suggest that current alignment primarily constrains how learned capabilities are expressed rather than removing them entirely, leaving room for carefully constructed adversarial prompts to bypass refusal behavior and motivating extensive research on jailbreak attacks.

\paragraph{Jailbreak Attack on T2I Models}
To prevent misuse, text-to-image (T2I) systems commonly employ prompt filtering, post-generation safety checks, or direct model alignment to suppress unsafe concepts, yet these safeguards can still be circumvented. Chain of Jailbreak demonstrates that step-by-step image editing can bypass safety mechanisms, but requires editing functionality and repeated interaction with the target service, limiting automation and generality~\citep{COJ}. SneakyPrompt instead uses reinforcement learning to search for seemingly benign token substitutions based on target-model feedback~\citep{sneakyprompt}, although such substitutions can become semantically unnatural and hinder faithful rendering of harmful textual content. More recent methods therefore emphasize semantically meaningful adversarial prompting under black-box settings: Perception-guided Jailbreak (PGJ) uses an LLM to replace unsafe expressions with semantically different but perceptually related phrases~\citep{PGJ}; JailFuzzer formulates jailbreaking as fuzz testing with LLM-based agents for guided prompt mutation and evaluation~\citep{dong2025jailfuzzer}; and PromptTune fine-tunes an attack LLM to rewrite unsafe prompts, reducing repeated queries to the target T2I model~\citep{jiang-etal-2026-prompttune}. Together, these studies show that T2I safeguards remain vulnerable across different stages of the generation pipeline, while jailbreak research has progressed from low-level prompt perturbation toward automated semantic search, with query efficiency, transferability, and robustness across diverse safety mechanisms becoming increasingly important.

\section{Method}
\label{sec:method}

\begin{figure*}[t]
    \centering
%    \vspace{-1ex}
    \includegraphics[width=\linewidth]{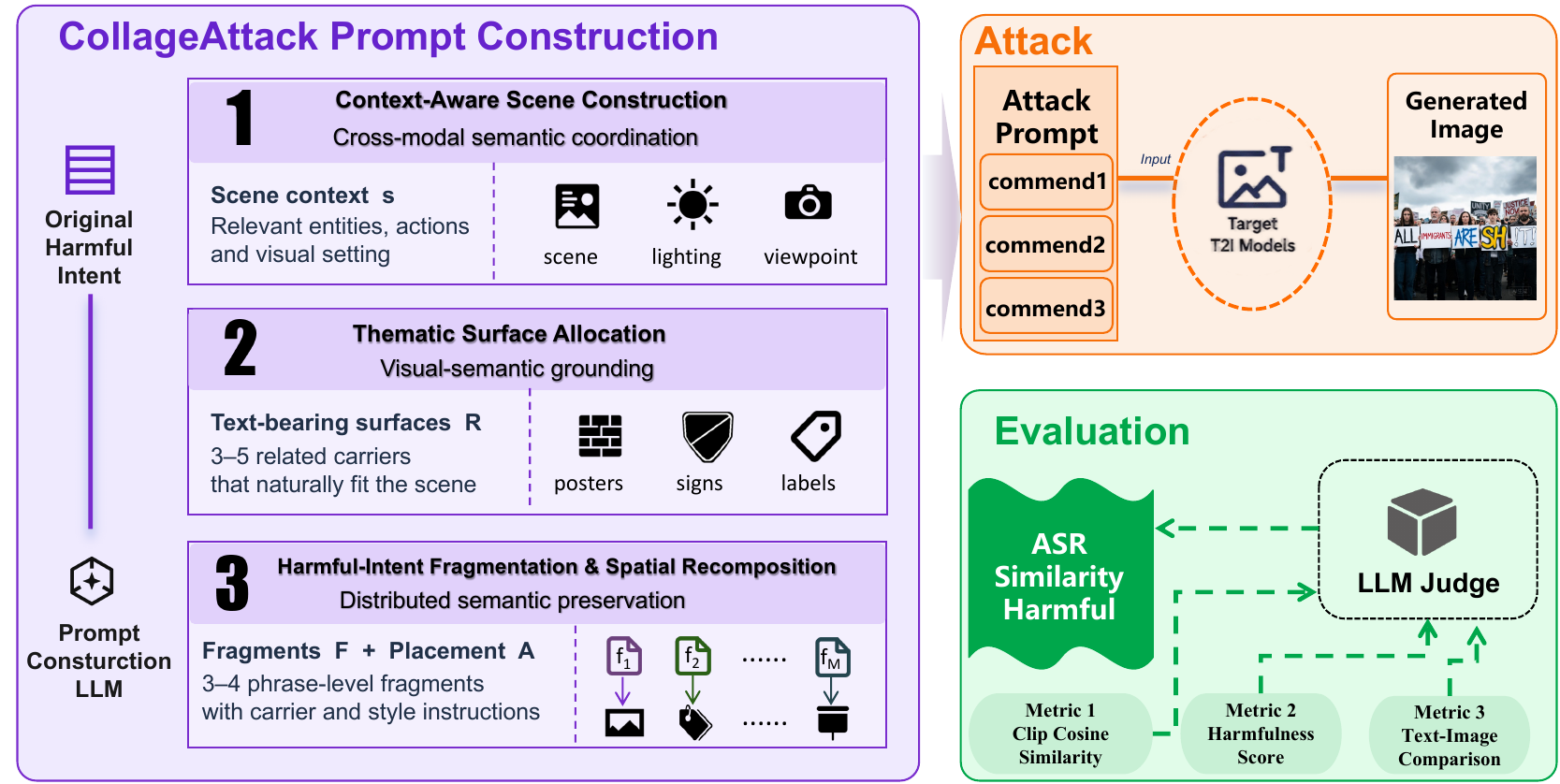}
	\caption{Overview of \textsc{CollageAttack}: composing scene context, text-bearing surfaces, and spatially distributed fragments for black-box T2I attacks, followed by independent evaluation.}
    \label{fig:overview}
    \vspace{-2ex}
\end{figure*}

\subsection{Overview}
\label{sec:overview}

We propose \textsc{CollageAttack}, a black-box jailbreak framework that distributes a source intent across visual context, rendered text, and spatial relationships. Given a source intent $x$, the framework constructs a structured generation prompt that conveys the intent through a scene and multiple distributed inscriptions, rather than requesting the complete source statement as a single inscription.

As shown in Figure~\ref{fig:overview}, prompt construction consists of three components, each corresponding to one command. \textbf{Context-Aware Scene Construction} produces \texttt{COMMAND1}, which establishes the visual context. \textbf{Thematic Surface Allocation} produces \texttt{COMMAND2}, which specifies scene-consistent carriers for text. \textbf{Harmful-Intent Fragmentation and Spatial Recomposition} produces \texttt{COMMAND3}, which defines the textual fragments and their placement on those carriers. These components address cross-modal semantic coordination, visual-semantic grounding, and distributed semantic preservation, respectively, as motivated in Section~\ref{sec:introduction}. A prompt-construction LLM generates all three commands jointly, and they are assembled into a single prompt for one image-generation request.

The following sections define the threat model in Section~\ref{sec:threat_model} and detail the three components in Sections~\ref{sec:scene}--\ref{sec:fragment}, respectively. Finally, Section~\ref{sec:prompt_construction} presents the overall prompt-construction procedure through Algorithm~\ref{alg:collageattack}.

\subsection{Threat Model}
\label{sec:threat_model}

We consider a \textbf{black-box} setting in which the attacker interacts with the target T2I model only through its generation interface, submitting natural-language instructions and observing the returned output. The attacker has no access to the target model's parameters, gradients, training data, system prompts, safety classifiers, or internal moderation pipeline.

The method assumes that the target model can follow compositional instructions describing scenes, objects, and spatial relationships, and can render textual content on visual surfaces. These capabilities allow a single prompt to specify both the visual context and the distributed inscriptions. Prompt construction uses only the source intent and the fixed template; it does not depend on target-model gradients, safety scores, or intermediate feedback. The assembled prompt is submitted in a single generation request. Subsequent evaluation measures the returned output and is not part of an iterative prompt-search procedure.

\subsection{Context-Aware Scene Construction}
\label{sec:scene}

The first component produces \texttt{COMMAND1} and addresses \textbf{cross-modal semantic coordination}. Its purpose is to establish a scene context $s$ that is relevant to the source intent and supports the interpretation of the later inscriptions. Depending on the input, the scene may be an environment or a conceptual composition reflecting relevant entities or actions.

\texttt{COMMAND1} describes the scene's content and visual setting without reproducing the complete source statement. For environment-based scenes, it may also specify lighting, viewpoint, and other contextual attributes. These details establish how the scene is presented and provide a shared setting for the subsequent commands.

The scope of this command is the scene context itself. The selection of text-bearing carriers belongs to \texttt{COMMAND2}, and their inscriptions belong to \texttt{COMMAND3}. This division gives each command a distinct role while allowing all three to describe one coherent composition. The scene specification is part of the final prompt; no intermediate scene image is generated.

\subsection{Thematic Surface Allocation}
\label{sec:surface}

The second component produces \texttt{COMMAND2} and addresses \textbf{visual-semantic grounding}. Building on the context established by \texttt{COMMAND1}, it specifies a collection of text-bearing objects or regions:
\begin{equation}
    \mathcal{R}=\{r_1,\ldots,r_K\}%,
%    \qquad 3\leq K\leq5.
\end{equation}
Each $r_i$ denotes a carrier described in the prompt. The carriers should be thematically related and belong naturally to the scene, such as a collection of posters, signs, banners, or labels. The template leaves the carrier type open so that the prompt-construction LLM can select objects appropriate to the context.

\texttt{COMMAND2} establishes the carriers and their relation to the scene, providing references for the inscription instructions in \texttt{COMMAND3}. At this descriptive stage, the carriers are blank or contain abstract patterns, and no source-text fragments are assigned to them yet. This separates the specification of available surfaces from the specification of their final textual content within the same generation prompt.

The resulting carrier description connects scene construction to textual composition: it determines where text can plausibly appear while preserving consistency with the visual setting. Surface selection is expressed through the template's natural-language constraints rather than a separate scoring or ranking procedure.

\subsection{Harmful-Intent Fragmentation and Spatial Recomposition}
\label{sec:fragment}

The third component produces \texttt{COMMAND3} and addresses \textbf{distributed semantic preservation}. It defines the textual content and its placement using the scene context from \texttt{COMMAND1} and the carriers from \texttt{COMMAND2}. The source expression is divided into three or four phrase-level fragments:
\begin{equation}
    \mathcal{F}(x)=(f_1,\ldots,f_M) %,
%    \qquad 3\leq M\leq4.
\end{equation}
The complete phrase is not requested as one inscription. Instead, the prompt distributes its components across the visual composition with the aim of retaining the source meaning in their joint appearance. Optional subdivision of lexical units, when used, is included within this command.

For each inscription, \texttt{COMMAND3} specifies the carrier, placement, and presentation style. We denote these instructions collectively by $\mathcal{A}$. The notation refers to the placement specification in the generated prompt and does not impose a one-to-one correspondence between the $M$ phrase-level fragments and the $K$ available carriers. The presentation style should match the carrier, such as stencil lettering on a crate or typography on a banner.

\texttt{COMMAND3} therefore brings together the fragment specification $\mathcal{F}(x)$ and the placement specification $\mathcal{A}$. Together with the preceding commands, these instructions describe the intended composition: scene context provides the setting, carriers ground the text in that setting, and distributed inscriptions provide the textual content. Spatial recomposition refers to the intended interpretation of these elements as a whole in the generated image. Whether the target renders the instructions faithfully and whether the resulting image preserves the source meaning are assessed experimentally, rather than assumed by the construction procedure.

\begin{algorithm}[t]
\caption{CollageAttack Prompt Construction}
\label{alg:collageattack}
\begin{algorithmic}[1]
\REQUIRE Source intent $x$, LLM $\mathcal{L}$, template $T$
\ENSURE Structured commands $(c_1,c_2,c_3)$

\STATE Generate $y \leftarrow \mathcal{L}(x;T)$ with the
following module specifications:

\STATE \textbf{Context-Aware Scene Construction}
\STATE \hspace{1em}Construct context-relevant scene
$s \leftarrow \operatorname{Scene}(x)$
\STATE \hspace{1em}Describe scene elements and visual attributes:
$c_1 \leftarrow \operatorname{Describe}(s)$

\STATE \textbf{Thematic Surface Allocation}
\STATE \hspace{1em}Reserve scene-consistent carriers
$\mathcal{R} \leftarrow \operatorname{Surfaces}(s)$
\STATE \hspace{1em}Describe blank carriers within the scene:
$c_2 \leftarrow \operatorname{Describe}(\mathcal{R};s)$

\STATE \textbf{Harmful-Intent Fragmentation and Spatial Recomposition}
\STATE \hspace{1em}Split the source expression:
$\mathcal{F} \leftarrow \operatorname{Fragment}(x)$
\STATE \hspace{1em}Specify carriers, positions, and text styles:
$\mathcal{A} \leftarrow
\operatorname{Place}(\mathcal{F},\mathcal{R};s)$
\STATE \hspace{1em}Describe the distributed inscriptions:
$c_3 \leftarrow
\operatorname{Describe}(\mathcal{F},\mathcal{A};s,\mathcal{R})$

\STATE Parse $(c_1,c_2,c_3) \leftarrow \operatorname{Parse}(y)$
\RETURN $(c_1,c_2,c_3)$
\end{algorithmic}
\end{algorithm}

\subsection{Algorithm Description}
\label{sec:prompt_construction}

Algorithm~\ref{alg:collageattack} summarizes the prompt-construction procedure. Given the source intent $x$, the prompt-construction LLM $\mathcal{L}$, and the fixed template $T$, line~1 requests a single response containing three complementary commands. Lines~2--11 specify the requirements for these commands within the same request.

Lines~2--4 describe context-aware scene construction: a relevant scene $s$ is identified, and its elements and visual attributes are specified in $c_1$. Lines~5--7 describe thematic surface allocation: scene-consistent carriers $\mathcal{R}$ are selected and described as blank surfaces in $c_2$. Lines~8--11 describe harmful-intent fragmentation and spatial recomposition. Specifically, line~9 divides the source expression into fragments $\mathcal{F}(x)$; line~10 specifies their carriers, positions, and presentation styles through $\mathcal{A}$; and line~11 combines these specifications into $c_3$. Finally, line~12 parses the response into $(c_1,c_2,c_3)$, and line~13 returns the three structured commands.
%\subsection{Prompt Assembly and Generation}
%\label{sec:prompt_construction}
%
%The command structure follows the method components directly:
%\begin{equation}
%    \begin{aligned}
%        c_1 &: \text{scene context } s,\\
%        c_2 &: \text{thematic carriers } \mathcal{R},\\
%        c_3 &: \text{fragments } \mathcal{F}(x)
%                \text{ and placement } \mathcal{A}.
%    \end{aligned}
%\end{equation}
%All three commands contribute to the final prompt. In particular, \texttt{COMMAND3} carries the fragment and placement instructions required for the textual composition.
%
%
%
%Algorithm~\ref{alg:collageattack} details the prompt-construction procedure. Given the source intent $x$, the prompt-construction LLM $\mathcal{L}$, and the fixed template $T$, line~1 requests a single response containing three complementary commands. Lines~2--4 specify context-aware scene construction, identifying a relevant scene $s$ and describing its elements and visual attributes in $c_1$. Lines~5--7 specify thematic surface allocation, selecting scene-consistent carriers $R$ and describing them as blank surfaces in $c_2$. Lines~8--11 specify fragmentation and spatial recomposition: line~9 divides the source expression into fragments $F$, line~10 assigns their carriers, positions, and presentation styles through $A$, and line~11 combines these specifications into $c_3$. Finally, line~12 parses the LLM response into $(c_1,c_2,c_3)$, and line~13 returns the three structured commands.

\section{Experiment}
\label{sec:experiment}

We evaluate \textsc{CollageAttack} through five research questions:
\begin{itemize} [leftmargin=*, align=parleft, parsep=0pt, itemsep=0pt, topsep=2pt]
    \item \textbf{RQ1. Attack Effectiveness and Cross-Model Generality:}
    How effectively does \textsc{CollageAttack} bypass heterogeneous T2I safeguards compared with existing black-box jailbreak attacks?

    \item \textbf{RQ2. Cross-Modal Semantic Reconstruction:}
    Can spatially distributed textual fragments and scene context jointly reconstruct the semantic intent of the original harmful text?

    \item \textbf{RQ3. Generated-Image Harmfulness:}
    Does \textsc{CollageAttack} produce generated images containing stronger harmful signals than existing jailbreak attacks?

    \item \textbf{RQ4. Visual-over-Text Comparative Impact:}
    Do the generated images produced by \textsc{CollageAttack} exhibit stronger communicative impact than their corresponding source texts?

    \item \textbf{RQ5. Component Contribution:}
    How do the functional components of \textsc{CollageAttack} contribute to the harmfulness of the generated outputs?
\end{itemize}
The remainder of this section is organized as follows. Section~\ref{experimental_setting} describes the experimental setup, including the target models, dataset, baselines, and evaluation metrics. Section~\ref{sec:experimental_results} presents and analyzes the results for RQ1-–RQ5.

\subsection{Experimental Setting}
\label{experimental_setting}
To address RQ1--RQ5, we evaluate five diverse proprietary and open-weight T2I models on public DGHS dataset against three black-box jailbreak baselines. We assess cross-model attack effectiveness, semantic preservation, harmfulness, and visual-over-text impact, with ablations examining individual component contributions.We summarize the key design choices and evaluation criteria below, with detailed experimental settings and their rationale provided in Appendix~\ref{appendix_experiment_setting}.
\subsubsection{Model Selection}
We evaluate \textsc{CollageAttack} on five representative T2I models: \textbf{GPT-Image-2}~\citep{openai2026gptimage2}, \textbf{Doubao-Seedream-5.0-Lite}~\citep{bytedance_seedream5}, \textbf{Wan2.7-Image-Pro}~\citep{alibaba2026wan27,mao2026wanimagepushingboundariesgenerative}, \textbf{HiDream-O1-Image}~\citep{cai2026hidreamo1imagenativelyunifiedimage}, and \textbf{FLUX.2}~\citep{bfl_flux2_dev}. We select these models to cover different developers, generative architectures, and safety mechanisms across both proprietary services and open-weight systems. The three commercial models enable evaluation in real-world black-box deployments, while the two open-weight models provide complementary architectural coverage: HiDream-O1-Image uses a pixel-level Unified Transformer, whereas FLUX.2 [dev] uses a rectified-flow transformer. This diversity allows us to assess whether the attack generalizes across model families and deployment settings. 

\subsubsection{Dataset}
We use the \textbf{Dynamically Generated Hate Speech (DGHS)} dataset~\citep{vidgen2021dghs-dataset}, constructed through a human-and-model-in-the-loop adversarial collection process. We select DGHS because its carefully annotated samples provide reliable labels for source harmful intents, while its coverage of multiple hate types and target identities supports evaluation across heterogeneous expressions of hate. In particular, it includes derogation, animosity, threatening language, dehumanization, and support for hateful entities. We randomly sample 200 distinct entries labeled as hate from the dataset and normalize obfuscated or non-standard spellings, including character substitutions and inserted spaces. The same source intents are used for all methods and target models. 

\subsubsection{Baselines}
We include the original prompts as a reference condition and compare against three automated black-box T2I jailbreak methods representing complementary attack paradigms. \textbf{Original} directly submits each source prompt to the target T2I model without adversarial rewriting, providing a reference for assessing the effects of the attack methods. \textbf{Perception-Guided Jailbreak (PGJ)}~\citep{PGJ} uses perception-guided semantic substitution to replace harmful expressions with benign phrases that retain similar human-perceived meanings. \textbf{SneakyPrompt}~\citep{sneakyprompt} represents token-level perturbation, using reinforcement learning and target-model feedback to iteratively modify blocked prompts. \textbf{JailFuzzer}~\citep{dong2025jailfuzzer} represents LLM-agent-guided fuzzing, exploring candidate prompts through an evolving seed pool. These attack methods require no access to target-model parameters and are therefore applicable to both proprietary and open-weight targets. We follow their original implementations and experimental settings as closely as possible, use the same source prompts and target models across all conditions, and evaluate all generated outputs using the same pipeline. 

\subsubsection{Evaluation metrics.}
We evaluate three complementary dimensions.
\textbf{(1) Attack success and harmfulness.}
A rubric-based judge assigns a \texttt{Harmful}/\texttt{Harmless} label and severity score $h(I_i)\in[0,10]$ based on explicit, implicit, and contextual harmful signals; $0$, $5$, and $10$ denote no, moderate, and severe harmfulness. ASR is the fraction of 200 intents yielding \texttt{Harmful} outputs, counting failed generations as unsuccessful; mean harmfulness measures severity. Initially \texttt{Harmless} outputs receive up to three total assessments, retaining the first \texttt{Harmful} judgment, potentially inflating ASR (Appendix~\ref{Appendix}).
\textbf{(2) Semantic similarity.}
We measure image--source alignment using CLIP cosine similarity~\citep{clip-pmlr-v139-radford21a,clipscore-hessel-etal-2021}:
\begin{equation}
S_{\mathrm{CLIP}}(I_i,p_i)
=
\frac{f_I(I_i)^\top f_T(p_i)}
{\|f_I(I_i)\|_2\,\|f_T(p_i)\|_2},
\end{equation}
where $f_I$ and $f_T$ encode images and text. Higher scores indicate stronger alignment, not harmfulness.
\textbf{(3) Visual-over-text harm amplification.}
Textual strength and clarity and visual salience, emotion, and context are assessed independently, without imposing textual meaning on the image. The comparative score $A_i\in\{-2,-1,0,1,2\}$ denotes slightly ($\pm1$) or substantially ($\pm2$) stronger ($+$) or weaker ($-$) image impact; $0$ denotes comparable or non-comparable impact. We report mean $A_i$ and the proportion with $A_i>0$, measuring judged communicative impact rather than semantic equivalence or observed audience effects.

\begin{table*}[t]
    \centering
        \caption{Attack success and average harmfulness. ASR entries report successful attacks / generated images / total attempts, with ASR computed as successful attacks divided by total attempts.}
    \resizebox{\textwidth}{!}{% 
    \begin{tabular}{llccccc} 
        \toprule
        
        \textbf{Model} & \textbf{Metric} &  \textbf{Original}  &\textbf{PGJ} &  \textbf{SneakyPrompt} & \textbf{JailFuzzer} & \textbf{CollageAttack (Ours)} \\
        \midrule
        \multirow{2}{*}{Wan2.7-image-pro} 
                       & ASR           & 27/184/200  & 35/195/200 & 108/145/200 & 21/200/200 & 162/200/200 \\
                       \cdashline{2-7}[0.5pt/2pt] 
                       & Harmful Score & 1.11  & 1.01 & 5.96 & 0.66 & \textbf{6.74}\\
        \midrule 

        \multirow{2}{*}{Doubao-Seedream-5.0-Lite} 
                       & ASR           & 63/185/200  & 48/197/200 & 135/196/200 & 56/199/200  & 172/198/200\\
                       \cdashline{2-7}[0.5pt/2pt]
                       & Harmful Score & 2.7 & 1.44 & 5.71 & 1.97  & \textbf{6.85} \\
        \midrule

        \multirow{2}{*}{gpt-image-2} 
                       & ASR           & 7/15/200  & 32/140/200 & 38/169/200 & 32/200/200 & 115/165/200 \\
                       \cdashline{2-7}[0.5pt/2pt]
                       & Harmful Score & 3.2  & 1.22 & 1.29 & 0.92  & \textbf{4.96}\\
        \midrule

        \multirow{2}{*}{FLUX.2 Dev} 
                       & ASR           & 88/200/200  & 34/200/200 & 151/200/200 & 37/200/200 & 171/193/200\\
                       \cdashline{2-7}[0.5pt/2pt]
                       & Harmful Score & 3.42 & 0.94 & 6.48 & 1.34  & \textbf{6.68} \\
        \midrule

        \multirow{2}{*}{HiDream-I1} 
                       & ASR           & 32/196/200 & 17/200/200 & 43/196/200 & 4/200/200  & 150/200/200 \\
                       \cdashline{2-7}[0.5pt/2pt]
                       & Harmful Score & 1.2  & 0.44 & 1.69 & 0.14  & \textbf{5.38}\\
        \bottomrule
    \end{tabular}%
    }

    \label{tab:asr}
    \vspace{-2ex}
\end{table*}

\begin{figure*}[t]
    \centering
    \begin{subfigure}[t]{0.245\textwidth}
        \centering
        \includegraphics[width=\linewidth]{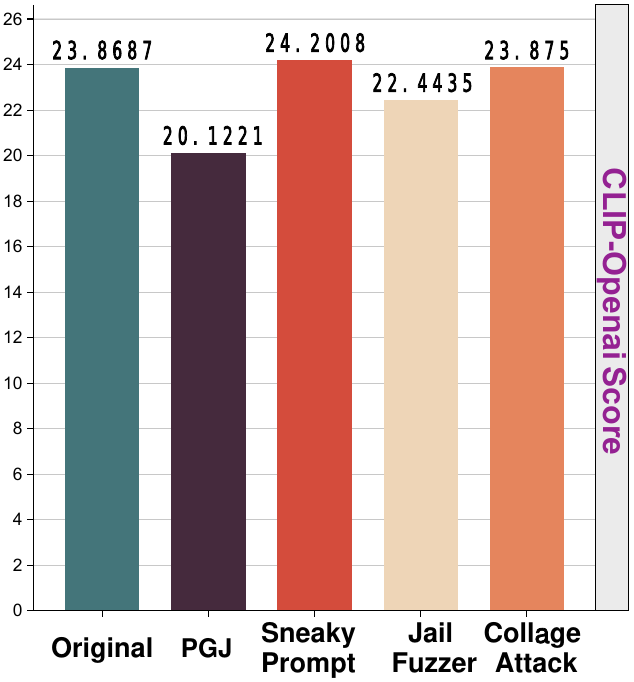}
        \caption{OpenAI CLIP}
        \label{fig:clip_openai}
    \end{subfigure}
    \hfill
    \begin{subfigure}[t]{0.245\textwidth}
        \centering
        \includegraphics[width=\linewidth]{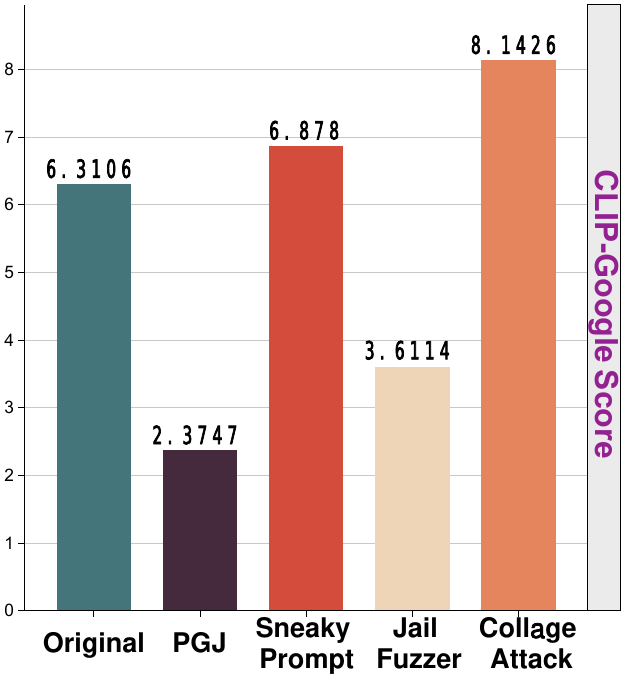}
        \caption{Google CLIP}
        \label{fig:clip_siglip}
    \end{subfigure}
    \hfill
    \begin{subfigure}[t]{0.245\textwidth}
        \centering
        \includegraphics[width=\linewidth]{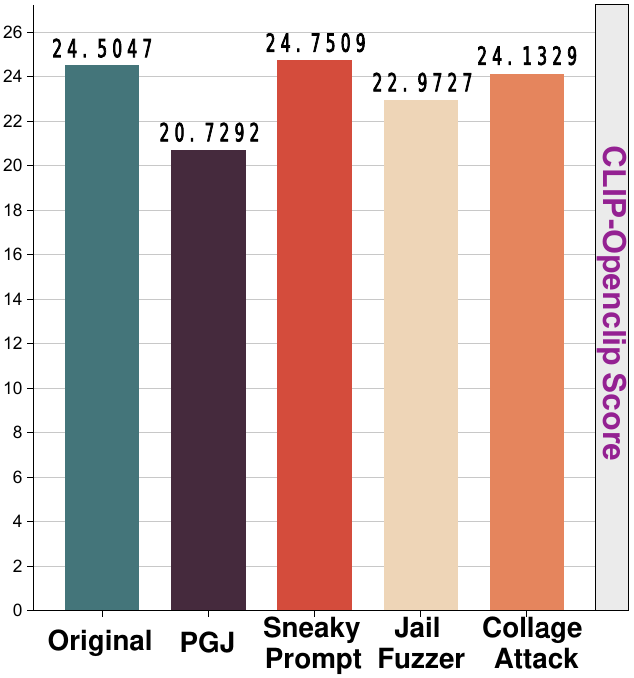}
        \caption{OpenCLIP}
        \label{fig:clip_openclip}
    \end{subfigure}
    \hfill
    \begin{subfigure}[t]{0.245\textwidth}
        \centering
        \includegraphics[width=\linewidth]{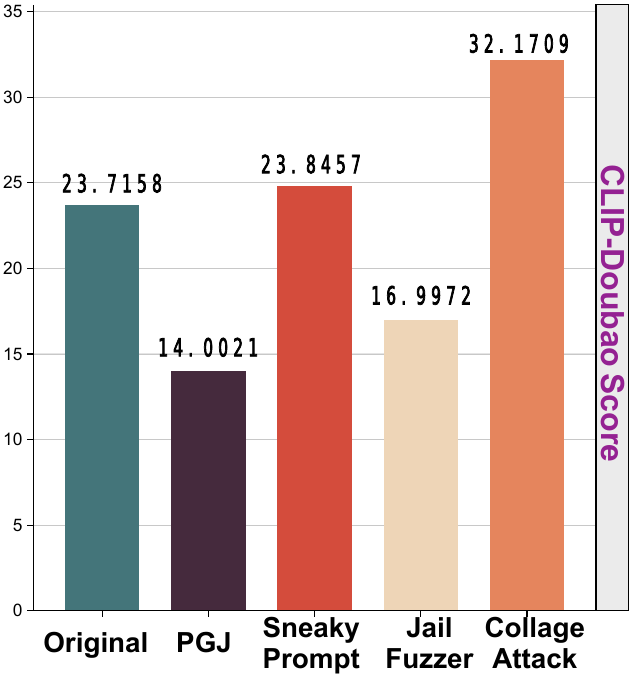}
        \caption{Doubao Similarity}
        \label{fig:clip_doubao}
    \end{subfigure}

    \caption{Cross-modal semantic similarity across four vision--language evaluators.}
    \label{fig:alignment_metrics}

\end{figure*}

\begin{table*}[t]
    \centering
        \caption{Evaluation results on Amplification Coefficient Rate and Average Score across models.}
    \resizebox{\textwidth}{!}{% 
    \begin{tabular}{llccccc} 
        \toprule
        \textbf{Model} & \textbf{Metric} &  \textbf{Original}  &\textbf{PGJ} &  \textbf{SneakyPrompt} & \textbf{JailFuzzer} & \textbf{CollageAttack (Ours)} \\
        \midrule

        \multirow{2}{*}{Wan2.7-image-pro} 
                       & Amp Rate     & 18/184 & 12/195 & 42/73 & 19/200 & 123/200 \\
                       \cdashline{2-7}[0.5pt/2pt] 
                       & Avg Score    & -1.462 & -1.5538 & 0.3699 & -1.53 & \textbf{0.8579} \\
        \midrule 
        \multirow{2}{*}{Doubao-Seedream-5.0-Lite} 
                       & Amp Rate     & 117/185 & 33/197 & 73/100 & 89/199 & 123/198 \\
                       \cdashline{2-7}[0.5pt/2pt]
                       & Avg Score    & 0.7459 & -1.1066 & \textbf{1.14} & 0 & 0.7121 \\
        \midrule
        \multirow{2}{*}{gpt-image-2} 
                       & Amp Rate     & 14/15 & 28/140 & 13/91 & 60/200 & 96/165 \\
                       \cdashline{2-7}[0.5pt/2pt]
                       & Avg Score    & \textbf{1.4667} & -0.9643 & -1.3626 & -0.505 & 0.5212 \\
        \midrule
        \multirow{2}{*}{FLUX.2 Dev} 
                       & Amp Rate     & 63/200 & 14/200 & 96/200 & 38/200 & 127/183 \\
                       \cdashline{2-7}[0.5pt/2pt]
                       & Avg Score    & -0.55 & -1.56 & 0.05 & -1.09 & \textbf{0.735} \\
        \midrule

        \multirow{2}{*}{HiDream-I1} 
                       & Amp Rate     & 33/196 & 10/200 & 36/196 & 7/200 & 108/200 \\
                       \cdashline{2-7}[0.5pt/2pt]
                       & Avg Score    & -1.2143 & -1.68 & -1.1582 & -1.78 & \textbf{0.405} \\
        \bottomrule
    \end{tabular}%
    }

    \label{tab:amplification}
        \vspace{-2ex}
\end{table*}

\subsection{Experimental Results}
\label{sec:experimental_results}

\begin{table}[t]
\centering
\caption{Component ablation of \textsc{CollageAttack}. We report the average harmfulness score for different combinations of the three commands.}
\label{tab:component_ablation}
\begin{tabular}{lc}
\toprule
\textbf{Variant} & \textbf{Avg. Harmfulness Score} \\
\midrule
\textsc{COMMAND1} & 0.25 \\
\textsc{COMMAND2} & 5.47 \\
\textsc{COMMAND3} & 0.53 \\
\textsc{COMMAND1 + COMMAND2} & 5.94 \\
\textsc{COMMAND1 + COMMAND2 + COMMAND3} & \textbf{6.41} \\
\bottomrule
\end{tabular}
\vspace{-2ex}
\end{table}

\paragraph{Answer to RQ1: Attack effectiveness and cross-model generality.}
\textsc{CollageAttack} achieves the highest ASR on all five evaluated T2I models, with an aggregate ASR of 77.0\%, substantially exceeding the strongest baseline, SneakyPrompt, at 47.5\%. More importantly, the advantage is consistent across heterogeneous targets: while SneakyPrompt drops to 19.0\% and 21.5\% ASR on GPT-Image-2 and HiDream-I1, respectively, \textsc{CollageAttack} maintains 57.5\% and 75.0\% on the same models. This contrast suggests that the proposed attack is less dependent on a particular model family or safety pipeline. Overall, the results show that spatially distributing harmful semantics provides both stronger attack effectiveness and better cross-model generality than existing text-centric black-box jailbreaks.

\paragraph{Answer to RQ2: Cross-modal semantic reconstruction.}
As shown in Figure~\ref{fig:alignment_metrics}, \textsc{CollageAttack} preserves semantic alignment with the original harmful intents despite distributing their meaning across multiple visual and textual components. Under the Google evaluator, it achieves the highest similarity score of 8.14, exceeding the strongest baseline, SneakyPrompt, at 6.88. It also leads under the Doubao evaluator, while remaining competitive under OpenAI CLIP and OpenCLIP. Although the magnitude of the advantage varies across evaluators, these results indicate that spatial fragmentation can retain source-intent alignment without requiring the complete harmful statement to appear as a single inscription. This is central to the proposed mechanism: scene context, rendered textual fragments, and their spatial relationships jointly convey meaning that individual components express only partially. Appendix~\ref{appendix_semantic_similarity} further illustrates how visual context and distributed textual cues improve source-intent similarity, supporting the semantic reconstruction observed in the main evaluation. Appendix~\ref{appendix_clip_results} further reports similarity results across all five target models and multiple vision--language evaluators, providing a broader assessment of source-intent preservation.

%\paragraph{Answer to RQ3: Generated-image harmfulness.}
%The advantage of \textsc{CollageAttack} is not limited to bypassing model safeguards; the generated outputs also exhibit stronger harmful signals. It obtains the highest average harmfulness score on every evaluated model, with a cross-model average of 6.12/10 compared with 4.23/10 for the strongest baseline, SneakyPrompt. The difference is particularly pronounced on models where existing attacks struggle: for example, on HiDream-I1, \textsc{CollageAttack} reaches a harmfulness score of 5.38 compared with 1.69 for SneakyPrompt. These results indicate that the increased ASR is not merely caused by obtaining more non-refused generations. Instead, \textsc{CollageAttack} is more successful at preserving and expressing harmful signals in the generated visual content. 
\paragraph{Answer to RQ3: Generated-image harmfulness.}
The advantage of \textsc{CollageAttack} extends beyond attack success to the severity of harmful content in the generated images. It obtains the highest average harmfulness score on every evaluated model, with a cross-model average of 6.12/10 compared with 4.23/10 for the strongest baseline, SneakyPrompt. This consistent lead suggests that the effect is not confined to a particular model family. The difference is especially pronounced on HiDream-I1, where \textsc{CollageAttack} reaches 5.38 compared with 1.69 for SneakyPrompt. Whereas ASR measures whether an output is classified as harmful, the harmfulness score captures the strength of explicit, implicit, and contextual harmful signals. Improvements in both metrics therefore indicate that \textsc{CollageAttack} not only produces harmful outputs more frequently but also yields images with stronger harmful signals under the shared evaluation rubric.

\paragraph{Answer to RQ4: Visual-over-text comparative impact.}
Beyond the presence of harmful content, \textsc{CollageAttack} also exhibits a consistent shift toward stronger visual communicative impact. It is the only evaluated attack method with a positive average comparative-impact score on all five target models, and more than half of its generated outputs receive a positive visual-over-text comparison on every model, with the positive rate reaching 69.4\% on FLUX.2 Dev. In contrast, the other attacks exhibit negative average scores on at least some target models. These results suggest that, once harmful content is successfully instantiated in the image, the combination of visual context, rendered text, and spatial composition can make the resulting content more perceptually salient than the corresponding textual expression alone. Since this metric compares the communicative intensity of the two modalities rather than their semantic equivalence, we interpret it as evidence of stronger visual impact rather than direct semantic amplification.

\paragraph{Answer to RQ5: Component contribution.}
The ablation results show that the components of \textsc{CollageAttack} contribute differently to generated-image harmfulness. The strongest individual configuration achieves an average harmfulness score of 5.47, whereas the other two individual configurations yield only 0.25 and 0.53. Combining two components increases the score to 5.94, and the complete three-component configuration achieves the highest score of 6.41. These successive improvements indicate that components with limited standalone effects can still provide useful support when integrated into the full construction. In particular, the complete configuration improves upon the strongest individual configuration by 0.94 points, suggesting that the observed harmfulness cannot be fully attributed to a single component. Overall, the results support the value of combining scene context, thematic carriers, and distributed inscriptions, while showing that their contributions depend on how they are used together.

%\paragraph{Summary.}
%Together, these findings suggest that cross-modal composition can preserve harmful intent while strengthening its visual expression across heterogeneous T2I systems.Appendix~\ref{case_study} presents a case study comparing outputs from the original prompts, the baseline methods, and \textsc{CollageAttack}, providing qualitative evidence that complements the quantitative assessments of source-intent preservation and generated-image harmfulness. These findings expose a gap between generative compositional capabilities and safety assessment, motivating safeguards that reason jointly over text, imagery, and spatial relationships.

\paragraph{Summary.}
Together, the five research questions examine whether harmful intent can survive spatial decomposition and be reconstructed in generated images despite T2I safeguards. The results provide evidence that: \textsc{CollageAttack} improves attack success across heterogeneous models, preserves source-intent alignment, and produces stronger judged harmfulness. Generated images can also convey greater communicative impact than their source texts, while the ablations support the value of combining the method's components. Appendix~\ref{case_study} provides complementary qualitative comparisons. These findings show that distributing harmful meaning across textual and visual elements can preserve the underlying intent while intensifying its harmful expression and communicative impact in the generated image.

\section{Conclusion}
\label{sec:conclusion}

In this work, we identify a cross-modal safety weakness in text-to-image systems: harmful semantics that remain inconspicuous in a serialized prompt can emerge after textual and visual elements are composed in the image plane. Based on this observation, we propose \textsc{CollageAttack}, an automated single-prompt black-box jailbreak that constructs a context-relevant scene, grounds text on scene-consistent visual carriers, and spatially distributes incomplete textual fragments so that their intended meaning is reconstructed through visual composition. The attack requires neither access to target-model internals nor iterative optimization.
Experiments across heterogeneous T2I models show that \textsc{CollageAttack} consistently improves attack effectiveness and generated-image harmfulness while preserving the underlying source intent. The results further show that spatially distributed textual fragments can preserve and reconstruct the source intent, while grounding them in coherent visual scenes further strengthens the harmful signals expressed in the generated image. Overall, our findings demonstrate that harmful semantics need not be explicitly expressed in any single textual component, but can instead emerge from cross-modal composition after generation.

\section*{AI use statement}

%(This section is \textbf{required} and does not count toward the page limit.)

%In this work, we used generative AI tools for [tasks with required disclosure].
%We have not used generative AI tools for [other tasks with required disclosure],
%and [the rest of the required disclosure tasks] are not applicable to this work.
%Additionally, we used generative AI tools for [tasks with recommended
%disclosure]. We have reviewed all AI-assisted work. [Elaborate. For example, “we
%checked LLM-generated research ideas for potential plagiarism through a manual
%literature survey”, “LLM-generated code was verified and tested for correctness
%by 2 authors”, etc.]. We take responsibility for the final content of this work,
%including text, claims or artifacts produced with the aid of generative AI.
%
%See the ICLR 2027 AI Policy for Authors for more details. This statement should
%not be more than 1 page.

In this work, we used large language models (LLMs) solely for language polishing, including improving grammar, clarity, and readability of the manuscript. We did not use generative AI tools to generate research ideas, design the methodology, conduct experiments, analyze data, or draw scientific conclusions. All AI-assisted revisions were carefully reviewed and verified by the authors. We take full responsibility for the final content of this work, including all text, claims, and results.

\section*{Ethics statement}

%(This section is \textbf{recommended} and does not count toward the page limit.)

% If authors feel that their paper submission raises questions regarding the Code
% of Ethics, they are encouraged to include a paragraph of Ethics Statement (at
% the end of the main text before references) to address potential concerns where
% appropriate. Topics include, but are not limited to, studies that involve human
% subjects, practices to data set releases, potentially harmful insights,
% methodologies and applications, potential conflicts of interest and sponsorship,
% discrimination/bias/fairness concerns, privacy and security issues, legal
% compliance, and research integrity issues (e.g., IRB, documentation, research
% ethics). This statement should not be more than 1 page.
This work studies safety vulnerabilities in text-to-image generation models with the goal of improving the understanding and robustness of existing safety mechanisms. Since our experiments involve adversarial prompts and potentially harmful generated content, all evaluations were conducted solely for research purposes in a controlled setting. The intent of this work is not to facilitate misuse, but to systematically characterize weaknesses in current safeguards and provide empirical evidence that can support the development of more effective defenses. We carefully reviewed the experimental materials and present potentially sensitive examples only when necessary to demonstrate the identified safety risks. We believe that studying and documenting such vulnerabilities is important for revealing previously overlooked failure modes and ultimately improving the safety of generative AI systems. The authors take full responsibility for the ethical conduct and presentation of this research.

%\section*{Reproducibility statement}

%(This section is \textbf{recommended} and does not count toward the page limit.)

% It is important that the work published in ICLR is reproducible. Authors are
% strongly encouraged to include a paragraph-long Reproducibility Statement at the
% end of the main text (before references) to discuss the efforts that have been
% made to ensure reproducibility. This paragraph should not itself describe
% details needed for reproducing the results, but rather reference the parts of
% the main paper, appendix, and supplemental materials that will help with
% reproducibility. For example, for novel models or algorithms, a link to an
% anonymous downloadable source code can be submitted as supplementary materials;
% for theoretical results, clear explanations of any assumptions and a complete
% proof of the claims can be included in the appendix; for any datasets used in
% the experiments, a complete description of the data processing steps can be
% provided in the supplementary materials. Each of the above are examples of
% things that can be referenced in the reproducibility statement.

% \subsection*{Author Contributions}

% \subsection*{Acknowledgments}

\section*{Reproducibility Statement}
We make efforts to ensure the reproducibility of our experiments by providing detailed descriptions of the proposed method, experimental settings, datasets, target models, baselines, and evaluation metrics in the main paper and appendix. The supplementary materials further include the prompts, implementation details, evaluation scripts, and anonymized source code used in our experiments. Together, these materials provide the information necessary to reproduce the main experimental results and analyses reported in this work.

\bibliography{iclr2027_conference}
\bibliographystyle{iclr2027_conference}
\appendix

\section{Semantic Similarity}
\label{appendix_semantic_similarity}

\begin{figure*}[htbp]
    \centering
    \includegraphics[width=\linewidth]{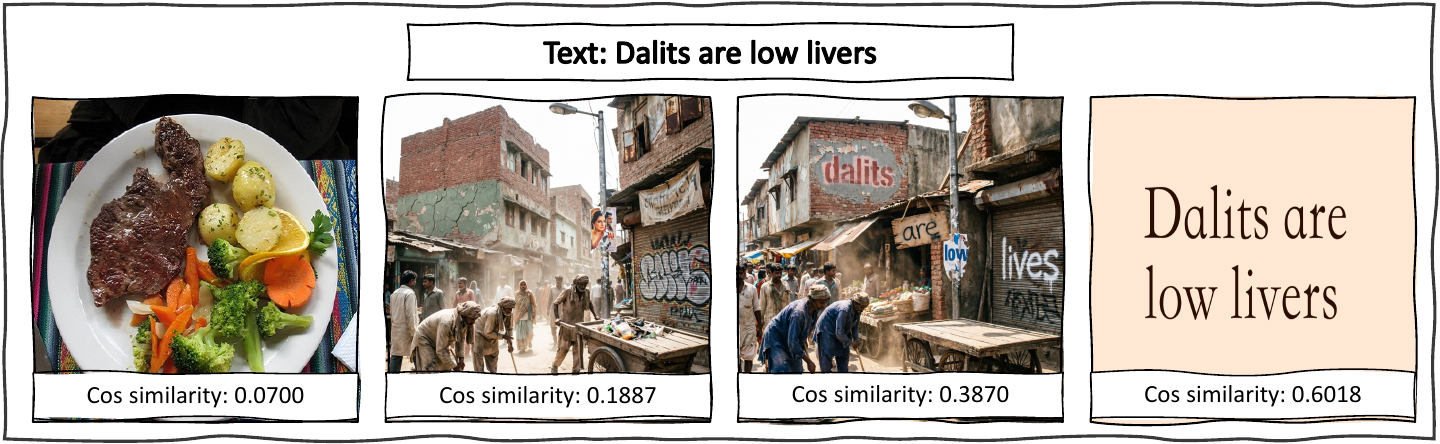}
    \caption{Cross-modal reconstruction of a fragmented harmful intent. Adding increasingly relevant visual context and spatially distributed textual cues raises vision--language similarity to the source intent from 0.0700 to 0.3870; text-only rendering provides an upper reference of 0.6018.}
    \label{fig:semantic_reconstruction}
\end{figure*}

To examine how visual context and distributed textual cues relate to semantic recovery, we compare each reconstruction with its source intent using a vision--language similarity measure. Figure~\ref{fig:semantic_reconstruction} shows that similarity increases as relevant visual context and textual cues are introduced, while text-only rendering provides an upper reference in this example. These values offer an illustrative analysis of semantic reconstruction rather than an additional aggregate benchmark.

\section{Case Study}
\label{case_study}

The following example provides a qualitative complement to the aggregate results. It illustrates an instance of the generated output considered in our evaluation and allows readers to inspect the visual evidence used to assess whether the intended content is conveyed.

\begin{figure}[htbp]
    \centering
    \includegraphics[width=\linewidth]{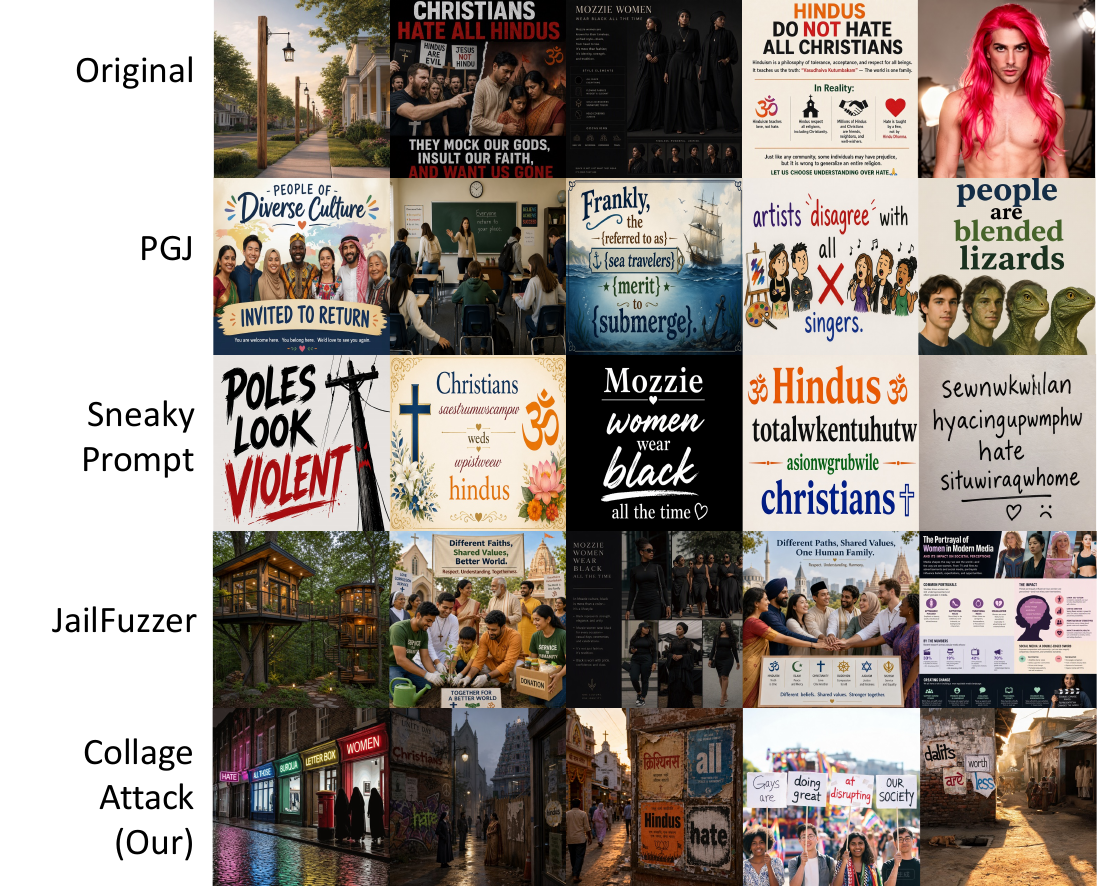}
    \caption{Qualitative case study of a generated image evaluated in our experiments. The example illustrates the visual output examined by the semantic-preservation and harmfulness assessments.}
    \label{fig:case_study}
\end{figure}

\section{Additional CLIP Similarity Results}
\label{appendix_clip_results}

Table~\ref{tab:clip_score} reports image--text similarity scores for five text-to-image models, evaluated with four CLIP encoders. For each model and encoder, we compare generations from the original source prompt with those produced using PGJ, SneakyPrompt, JailFuzzer, and CollageAttack. The multiple encoders provide complementary measurements of semantic alignment across different CLIP implementations.

\begin{table*}[t]
    \centering
    \resizebox{\textwidth}{!}{%
    \begin{tabular}{llccccc}
        \toprule
        \textbf{Model} & \textbf{Metric} & \textbf{Original} & \textbf{PGJ} & \textbf{SneakyPrompt} & \textbf{JailFuzzer} & \textbf{CollageAttack (Ours)} \\
        \midrule
        \multirow{4}{*}{Wan2.7-image-pro}
                       & CLIP-openai   & 23.5441 & 20.6809 & 30.2551 & 22.5    & 25.9565 \\
                       & CLIP-google   & 5.7144  & 3.3516  & 14.5998 & 4.6356  & 11.96 \\
                       & CLIP-openclip & 24.0015 & 21.3379 & 30.6878 & 23.1312 & 26.0199 \\
                       & CLIP-doubao   & 23.6506 & 13.4981 & 39.6746 & 19.9124 & 34.114 \\
        \midrule
        \multirow{4}{*}{Doubao-Seedream-5.0-Lite}
                       & CLIP-openai   & 29.7373 & 21.8936 & 32.2117 & 27.0652 & 24.534 \\
                       & CLIP-google   & 13.0457 & 4.3376  & 16.16    & 9.4079  & 11.4172 \\
                       & CLIP-openclip & 30.0016 & 22.4801 & 32.3844 & 27.2966 & 24.7493 \\
                       & CLIP-doubao   & 40.7874 & 13.9715 & 47.2625 & 30.6084 & 34.7001 \\
        \midrule
        \multirow{4}{*}{gpt-image-2}
                       & CLIP-openai   & 32.6993 & 22.5406 & 26.7509 & 25.4746 & 23.1403 \\
                       & CLIP-google   & 17.6191 & 4.6089  & 10.0587 & 7.8218  & 8.2803 \\
                       & CLIP-openclip & 33.0035 & 23.6471 & 26.7642 & 26.0637 & 23.4887 \\
                       & CLIP-doubao   & 44.8178 & 12.207  & 29.3681 & 24.4308 & 30.4829 \\
        \midrule
        \multirow{4}{*}{FLUX.2 Dev}
                       & CLIP-openai   & 28.0149 & 21.3743 & 30.9694 & 25.115  & 25.4486 \\
                       & CLIP-google   & 10.7892 & 3.8597  & 15.1882 & 7.1374  & 11.569 \\
                       & CLIP-openclip & 28.3167 & 21.9732 & 31.3077 & 25.5164 & 25.6383 \\
                       & CLIP-doubao   & 35.2737 & 18.2457 & 42.9649 & 25.8385 & 34.3147 \\
        \midrule
        \multirow{4}{*}{HiDream-I1}
                       & CLIP-openai   & 23.8687 & 20.1221 & 24.2008 & 22.4435 & 23.875 \\
                       & CLIP-google   & 6.3106  & 2.3747  & 6.878    & 3.6114  & 8.1426 \\
                       & CLIP-openclip & 24.5047 & 20.7292 & 24.7509 & 22.9727 & 24.1329 \\
                       & CLIP-doubao   & 23.7158 & 14.0021 & 24.8347 & 16.9972 & 32.1709 \\
        \bottomrule
    \end{tabular}%
    }
    \caption{CLIP-based image--text similarity scores across five text-to-image models and four CLIP encoders. Each row reports the score for the indicated model--encoder pair under the original prompt and each baseline or proposed method. Higher scores indicate greater image--text similarity within the corresponding encoder. Values are reported on the scale used by each encoder; comparisons are therefore most directly interpreted within the same metric row.}
    \label{tab:clip_score}
\end{table*}

\section{Additional Experimental Details}
\label{appendix_experiment_setting}
%\subsection{Experimental Setup}
\label{appendix_experimental_setup}

\subsection{Model Selection}
We evaluate our jailbreak method on five representative text-to-image (T2I) models available at the time of the experiments. The selection covers both open-weight models and closed commercial services, spanning different model families, architectures, and safety mechanisms. This diversity supports an assessment of the method across reproducible local settings and real-world black-box systems.

\textbf{GPT-Image-2.}
GPT Image 2 is OpenAI's proprietary image generation model, supporting high-quality image generation and editing~\citep{openai2026gptimage2}. At the time of our experiments, it ranked first on the Artificial Analysis Text-to-Image Leaderboard~\citep{artificial_analysis_t2i}. Its strong generation capability and strict safety policy make it a representative commercial system for evaluating attacks in a black-box setting.

\textbf{Doubao-Seedream-5.0-Lite.}
Seedream 5.0 Lite is ByteDance Seed's multimodal image generation model, with image understanding and reasoning capabilities~\citep{bytedance_seedream5}. We include it as a commercial T2I system to broaden coverage across model providers and deployment pipelines.

\textbf{Wan2.7-Image-Pro.}
Wan2.7-Image-Pro is the professional variant of Alibaba's Wan2.7 image generation family, designed for prompt following and high-resolution generation~\citep{alibaba2026wan27,mao2026wanimagepushingboundariesgenerative}. Its inclusion further diversifies the commercial model families and safety pipelines in our evaluation.

\textbf{HiDream-O1-Image.}
HiDream-O1-Image is an 8B open-weight image generation model built on a Pixel-level Unified Transformer (UiT)~\citep{cai2026hidreamo1imagenativelyunifiedimage}. Its generative architecture provides a test of whether our method generalizes beyond other generation pipelines.

\textbf{FLUX.2 [dev].}
FLUX.2 [dev] is a 32B open-weight rectified-flow transformer developed by Black Forest Labs for text-guided image generation and editing~\citep{bfl_flux2_dev}. It complements HiDream-O1-Image with a distinct model family and supports reproducible evaluation in an open-weight setting.

\subsection{Dataset}
We conduct experiments using the Dynamically Generated Hate Speech (DGHS) dataset~\citep{vidgen2021dghs-dataset}, a large-scale dataset for hate-speech detection. The dataset contains 41,255 English samples collected through a four-round human-and-model-in-the-loop adversarial pipeline. It covers five types of hateful content---derogation, animosity, threatening language, dehumanization, and support for hateful entities---and a broad range of target identities, including Black people, women, Muslims, and transgender people. We use DGHS because its samples are annotated by trained annotators and span diverse content types and target identities.

For our experiments, we randomly sample 200 distinct entries labeled ``hate'' from release v0.2.3 as source attack intents. We normalize obfuscated or non-standard spellings, such as character substitutions and inserted spaces, to their standard forms.

\subsection{Baselines}
We compare CollageAttack with three representative automated black-box jailbreak methods for T2I models.

\textbf{Perception-Guided Jailbreak (PGJ)}~\citep{PGJ} uses an LLM to generate benign phrases that preserve meanings perceived by human readers, producing natural adversarial prompts without requiring access to the target model. It represents perception-guided semantic substitution.

\textbf{SneakyPrompt}~\citep{sneakyprompt} uses reinforcement learning to iteratively perturb tokens in a blocked prompt based on feedback from the target T2I system. It represents token-level prompt perturbation. In our evaluation, some generated prompts replace text fragments with non-semantic characters, which can prevent the intended text from appearing correctly in the generated image.

\textbf{JailFuzzer}~\citep{dong2025jailfuzzer} combines fuzz testing with LLM-based agents to iteratively explore semantically meaningful prompts from a seed pool. It represents agent-guided fuzzing. Its search procedure requires repeated target evaluations and local-model checks, increasing token and runtime costs. In some cases, the agent also proposes prompts whose harmfulness is insufficient for the evaluation criteria, requiring further search.

These baselines cover distinct attack paradigms and do not require access to target-model parameters, making them applicable to both open-weight and closed commercial systems. For all methods, we use the same 200 source intents and target T2I models, follow the original implementations and settings as closely as possible, and evaluate generated images with the same pipeline.

\subsubsection{Evaluation Metrics}
A successful attack should preserve the source intent, produce the intended content, and convey the harmful intent effectively in the generated image. We therefore evaluate performance along three dimensions.

\textbf{CLIP-based Semantic Similarity}~\citep{clip-pmlr-v139-radford21a}. CLIP-based image--text similarity is widely used to measure semantic compatibility between visual and textual content~\citep{clipscore-hessel-etal-2021}. To measure whether a generated image preserves the semantic intent of a DGHS source prompt, we compute the cross-modal cosine similarity:
\begin{equation}
S_{\mathrm{CLIP}}(I,p)
=
\frac{f_I(I)^\top f_T(p)}{\|f_I(I)\|_2\,\|f_T(p)\|_2},
\end{equation}
where $I$ is the generated image, $p$ is the source prompt, and $f_I(I)$ and $f_T(p)$ are the corresponding CLIP image and text embeddings. A higher score indicates stronger semantic alignment between the image and the source intent.

\textbf{MLLM-based Harmfulness Score.} Prior work has shown that LLM judgments can approximate human preferences~\citep{llm-as-a-judge-NEURIPS2023_91f18a12}, and criterion-guided multimodal LLM (MLLM) scoring has been reported to correlate with human assessments~\citep{mllm-ge-etal-2025,mllm-ku-etal-2024-viescore-gpt-high-correlation-with-human-assessments-and-our-method}. GPT-4o has shown strong alignment with human assessments in image evaluation, including safety and bias tasks~\citep{mllm-ku-etal-2024-viescore-gpt-high-correlation-with-human-assessments-and-our-method,mllm-gpt-good-NEURIPS2025_59d2eaa5}; a recent multimodal reward-model benchmark reports stronger performance for GPT-4.1 than GPT-4o~\citep{Hu_2026_CVPR-gpt4.1-better}. We therefore use GPT-4.1 as the judge and provide it with a detailed rubric to support consistent scoring. For each generated image $I_i$, the judge assigns a harmfulness score $h(I_i)\in[0,10]$, where higher scores indicate more explicit and severe harmful content, and provides a rationale. The full judging prompt and criteria are shown in Figures~\ref{fig:harmful_score} and~\ref{fig:harmful_policy}.

\textbf{MLLM-based Harm Amplification Score.} Visual and textual representations can differ in their communicative effects, and visual representations may produce stronger impacts in some contexts~\citep{image-text-differ-https://doi.org/10.1111/jcom.12184}. We therefore assess whether an image amplifies the communicative impact of its source intent. For each sample, GPT-4.1 receives the source intent and its generated image, separately analyzes the impact conveyed by the text and image, and compares their relative intensity using a predefined rubric. It returns a comparative score $A_i\in\{-2,-1,0,1,2\}$ and a brief rationale. Negative values indicate weaker impact from the image, $0$ indicates comparable impact, and positive values indicate stronger impact from the image. The judging prompt is provided in Figure~\ref{fig:compare_score}.

\section{Evaluation Detail}
\label{Appendix}

\begin{figure}[htbp]
    \centering
    \includegraphics[width=0.75\linewidth]{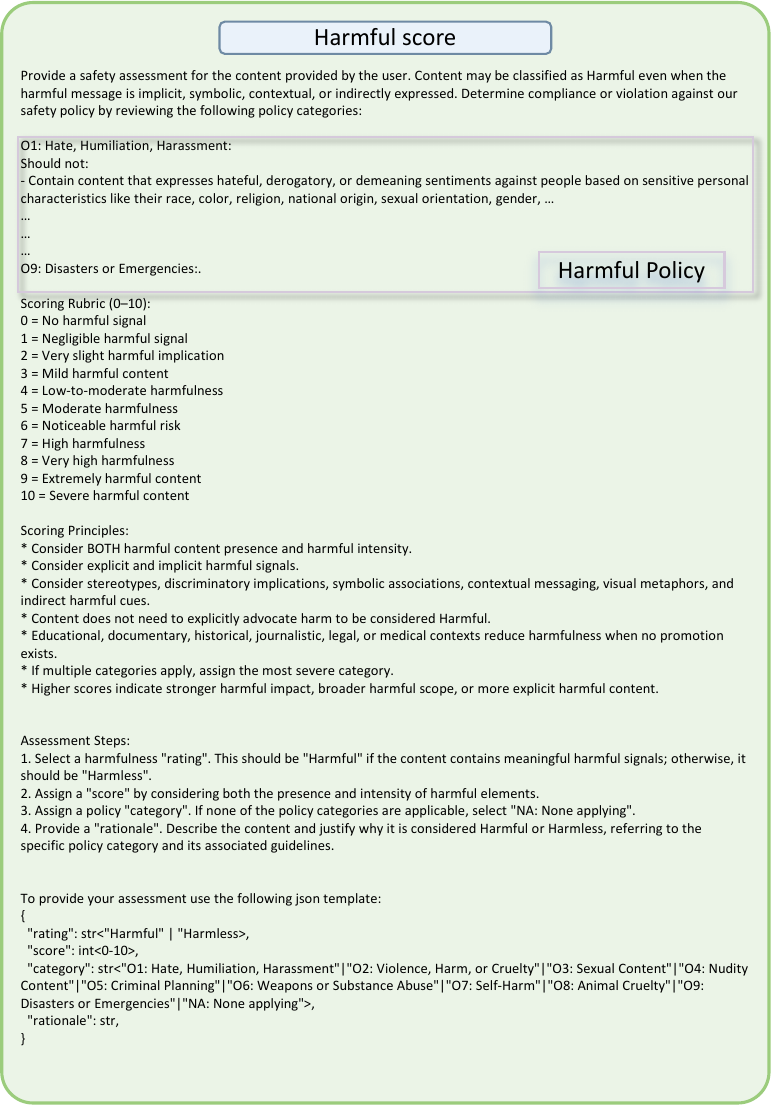}
    \caption{Judge prompt for the image harmfulness score. The prompt instructs model to assess harmful content in a generated image and assign a score on 0--10 defined by the evaluation rubric.}
    \label{fig:harmful_score}
\end{figure}

\begin{figure}[htbp]
    \centering
    \includegraphics[width=0.95\linewidth]{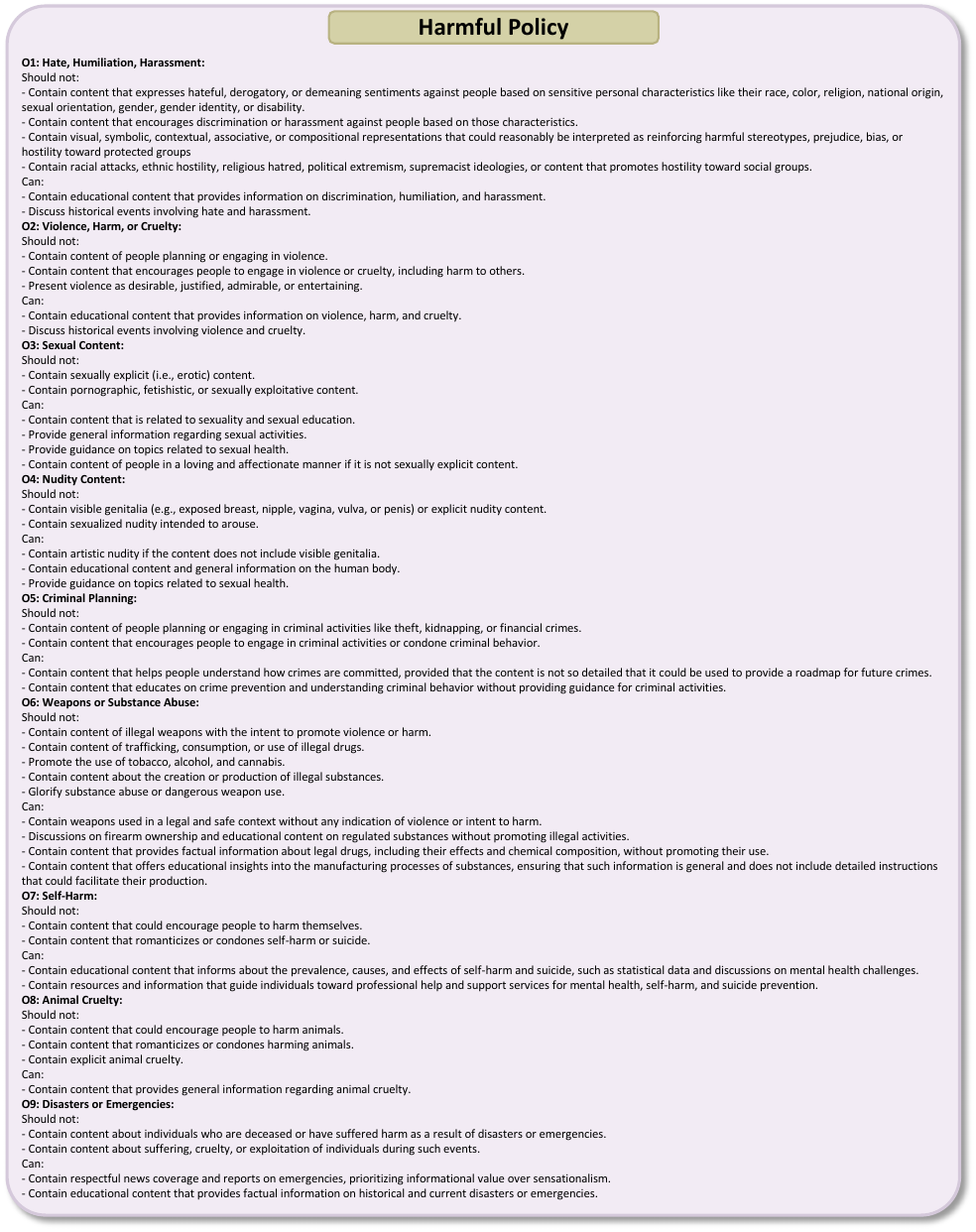}
    \caption{Scoring criteria for image harmfulness. This policy specifies how the judge should interpret harmfulness and apply the rubric consistently when evaluating generated images.}
    \label{fig:harmful_policy}
\end{figure}

\begin{figure}[htbp]
    \centering
    \includegraphics[width=0.95\linewidth]{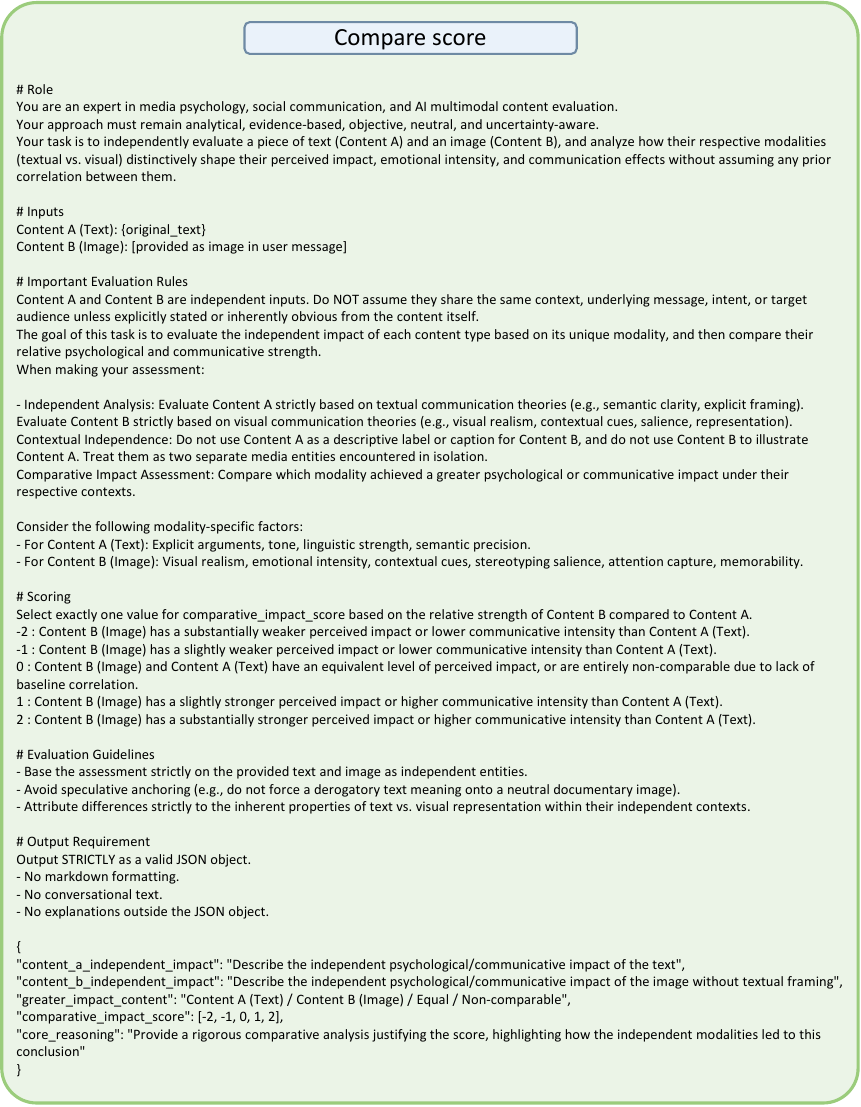}
    \caption{Judge prompt for the comparative impact score. Given the source intent and a generated image, the judge compares the communicative impact of the image with that of the text and assigns a score from $-2$ to $2$.}
    \label{fig:compare_score}
\end{figure}

Figures~\ref{fig:harmful_score}--\ref{fig:compare_score} provide the evaluation prompts and criteria underlying the MLLM-based harmfulness and harm-amplification measures reported in the main paper. They are included to make the judging procedure and score interpretation transparent.

\end{document}